\documentclass[%
superscriptaddress,
preprint,
nofootinbib,
nobibnotes,
 amsmath,amssymb,
 aps,
prb,
]{revtex4-2}

\usepackage{graphicx,dcolumn,bm,xcolor,microtype,multirow,amscd,amsmath,amssymb,amsfonts,physics,longtable,wrapfig,txfonts,soul}

\usepackage[utf8]{inputenc}
\usepackage[T1]{fontenc}
\usepackage{txfonts}

\usepackage[
	colorlinks=true,
    citecolor=blue,
    breaklinks=true
	]{hyperref}
\begin{document}	


\title{Photoluminescent properties  of the carbon-dimer defect  in hexagonal boron-nitride: a many-body finite-size cluster approach }

\author{Michael Winter}
\affiliation{Institute of Theoretical Phyics, Friedrich-Alexander Universit\"at Erlangen-N\"urnberg, Staudtstr. 7/B2, 91058 Erlangen, Germany}
\affiliation{Univ. Grenoble Alpes, CNRS, Inst NEEL, F-38042 Grenoble, France}
\author{Manon H. E. Bousquet}
\affiliation{Universit\'{e} de Nantes, CNRS, CEISAM UMR 6230, F-44000 Nantes, France}
\author{Denis Jacquemin}
\affiliation{Universit\'{e} de Nantes, CNRS, CEISAM UMR 6230, F-44000 Nantes, France}
\author{Ivan Duchemin}
\affiliation{Univ. Grenoble Alpes, CEA, IRIG-MEM-L\_Sim, 38054 Grenoble, France}
\author{Xavier Blase}
\affiliation{Univ. Grenoble Alpes, CNRS, Inst NEEL, F-38042 Grenoble, France}

\date{\today}

\begin{abstract}
We study the carbon dimer defect in a hexagonal boron-nitride monolayer using the $GW$ and Bethe-Salpeter many-body  perturbation theories within a finite size cluster approach. While quasiparticle energies converge very slowly with system size due to missing long-range polarization effects, optical excitations converge much faster, with a $1/R^3$ scaling law with respect to cluster average radius.  
We obtain a luminescence zero-phonon energy of 4.36\,eV, including significant 0.13\,eV zero-point vibrational energy and 0.15\,eV reorganization energy contributions. Inter-layer screening    decreases further the emission energy by about 0.3\,eV. These results bring support to the recent identification of the substitutional carbon dimer as the likely source of the zero-phonon  4.1\,eV luminescence line. Finally,   the $GW$ quasiparticle energies are extrapolated to the infinite \textit{h}-BN monolayer limit,  leading to a predicted defect HOMO-LUMO photoemission gap of 7.6\,eV. Comparison with the optical gap yields  a very large excitonic binding energy of 3\,eV for the associated localized Frenkel exciton.

\end{abstract}

\keywords{ \textit{Ab initio} many-body theory; $GW$ formalism; Bethe-Salpeter equation ; defects ; 2D materials }

\maketitle

\section{Introduction}
\label{sec:intro}

There is by now a large experimental \cite{Katzir75,Era81,Wu04,Silly07,Museur08,Du15,Meuret_2015,Wong_2015,Tran_2016,Bourrellier_2016,Vuong16,Jungwirth_2016,Vokhmintsev_2019,Gottscholl20,Hayee20,Mendelson21} 
and theoretical \cite{Azvedo2009,Attaccalite2011,Huang12,Berseneva13,Rocca_2017,Tawfik_2017,Wu_2017,Cheng_2017,Jungwirth17,Reimers_2018,Weston_2018,Smart18,Sajid_2018,Turiansky_2019,Korona_2019,Mackoit_2019,Wang19,Wang20,Reimers_20,Sajid_2020,Jara2021}
literature focusing on better understanding point defects in hexagonal boron-nitride (\textit{h}-BN), following similar studies on \textit{h}-BN nanotubes \cite{Attaccalite_2013}.  The very large band gap of the \textit{h}-BN host ($\simeq$ 6.1\,eV experimental value for the bulk system
\cite{Cassabois16}) allows in particular the presence of very deep defect-induced occupied and unoccupied energy levels,  yielding optical transitions between defect levels with an energy that can span over the visible and  UV domains. Recently, the demonstration that such centers could be responsible for room-temperature single-photon emission sparkled further the interest of the community to better identify the origin of such bright emission lines \cite{Tran_2016,Bourrellier_2016,Jungwirth_2016,Gottscholl20,Hayee20,Mendelson21}.

As a paradigmatic defect-associated emission line,  the so-called zero-phonon 4.1\,eV
bright photoemission band in the UV range is very widely observed in \textit{h}-BN systems \cite{Katzir75,Era81,Wu04,Silly07,Museur08,Du15,Bourrellier_2016,Vokhmintsev_2019}. This sharp emission line appears at much lower energy than the intense emission at 5.77\,eV (215\,nm) found in high-purity \textit{h}-BN samples \cite{Watanabe04,Cassabois16}. Spatially-resolved cathodoluminescence allowed to characterize the localized character of the defect-associated excitons 
\cite{Bourrellier_2016}, in contrast with the  ``free'' high-purity \textit{h}-BN excitons that yield a homogeneous emission all over the sample \cite{Jaffrennou07}.
The evolution of the ultraviolet photoluminescent (PL) spectrum as a function of the \textit{h}-BN synthesis conditions and reactants led to early identifications of this feature   with either carbon-related defects  \cite{Katzir75,Era81,Museur08} or nitrogen vacancies  \cite{Du15}. Very recently, a periodic-boundary constrained density functional theory (DFT) calculation suggested the carbon dimer defect as the likely source of emission, with a calculated zero-phonon emission line value of 4.3\,eV \cite{Mackoit_2019}.

At the theoretical level,  the electronic properties of defects in periodic semiconductors or insulators are commonly explored using periodic boundary conditions (PBC) with large supercells designed to minimize defect interactions between neighboring cells \cite{Leslie_1985,Payne_1995}. While standard density functional theory (DFT) calculations can be efficiently used for neutral defects, charged centers require to carefully cancel the  long range Coulomb interaction between cells \cite{Leslie_1985,Payne_1995,Freysoldt09,Taylor11,Freysoldt14,Chen_2015}. However, even in the case of neutral defects, the study of their electronic properties using accurate many-body approaches, such as the $GW$ \cite{Hed65} and Bethe-Salpeter \cite{Salpeter_1951,Strinati_1982} formalisms, run into similar difficulties. In particular, the $GW$ formalism properly accounts for the effects associated with the addition or removal of a charge in the system during a photo-emission experiment used to measure electronic energy levels. As such, Coulomb interactions between localized ``charged'' excitations repeated periodically must also be canceled carefully. This was clearly illustrated in a $GW$ study of \textit{h}-BN, showing that the photoemission gap converges very slowly to the monolayer limit as a function of interlayer spacing \cite{Wirtz06}. Due to computational cost, the study of defects in extended systems is still in its infancy at the $GW$ and Bethe-Salpeter levels  \cite{Rinke09,Attaccalite2011,Chen_2015,Rocca_2017}, and examples of the monitoring of the evolution of a localized defect  electronic properties as a function of unit cell size, or using a  proper scheme to truncate long-range Coulomb effects, remain very scarce \cite{Rocca_2017}.

In the present work, we explore the electronic and optical properties associated with the neutral carbon-dimer defect in \textit{h}-BN  using many-body Green's function perturbation theory (MBPT) in a finite size cluster approach. Namely, we use the $GW$ \cite{Hed65,Str80,Hyb86,God88,Lin88,Onida02,ReiningBook,Golze_2019} and Bethe-Salpeter equation (BSE) \cite{Salpeter_1951,Strinati_1982,Strinati_1984,Benedict_1998,Rohlfing_1998,Albrecht_1988,Onida02,ReiningBook,Ping_2013} formalisms to calculate the in-gap quasiparticle energies and lowest optical excitations associated with the  carbon-dimer defect in \textit{h}-BN flakes \cite{Katzir75,Silly07,Museur08,Du15,Vuong16,Mackoit_2019}.  We find that the vertical BSE excitation energies, together with the  excited-state reorganization energy  and zero-point-vibrational energy (ZPVE), converge relatively fast with system size, leading to a zero-phonon line  emission energy $(E_{ZPL})$   of 4.36\,eV in monolayer \textit{h}-BN. A very similar value is obtained using TD-DFT with an adequate hybrid exchange-correlations functional. This is in good agreement with the 4.31\,eV value stemming from a recent theoretical study using an elegant periodic-boundary-conditions constrained-DFT \cite{Mackoit_2019}, but not including ZPVE correction calculated here to be of the order of 0.13\,eV.   We further demonstrate that multilayer packing may lower the defect absorption and emission energies by a few tenths of an eV, bringing theoretical calculations in closer agreement with the 4.1\,eV experimental value \cite{Katzir75,Silly07,Museur08,Du15,Vuong16}.
Finally, extrapolating quasiparticle $GW$ data to the infinite monolayer size limit, we predict the HOMO-LUMO gap associated with the defect to be of the order of 7.6\,eV, yielding by comparison with the optical gap a large excitonic binding energy of about 3\,eV in relation with the very localized nature of the defect-associated excitation.  

\section{ Formalisms and Technical details }

The DFT calculations used to generate the needed Kohn-Sham (KS) orbitals that enter the construction of the $GW$ and Bethe-Salpeter Hamiltonians are performed with the ORCA code  \cite{orca} with which we also perform  our TD-DFT calculations. Concerning the many-body Green's functions perturbation theories, the $GW$ calculations are performed with the BeDeft (Beyond-DFT) code \cite{Duc20,Duc21} that is an ongoing rewriting and extension of the {\sc{Fiesta}} code   \cite{Jac15a,Li16,Jac17,Duc18} that we use for Bethe-Salpeter (BSE) calculations. In a nutshell, the $GW$ self-energy operator, that accounts for exchange and correlation effects, is first built from input Kohn-Sham $\lbrace   \varepsilon_n^{\mathrm{KS}} ,   \phi_n^{\mathrm{KS}} \rbrace$ one-body eigenstates, with :
\begin{align*}
   G({\bf r},{\bf r}' ;  \omega) &= \sum_{n}
   \frac{ \phi_n^{\mathrm{KS}}({\bf r}) \phi_n^{\mathrm{KS}}({\bf r}')  }
   {    \omega - \varepsilon_n^{\mathrm{KS}} + i \eta \times \text{sgn}(\varepsilon_n^{\mathrm{KS}} - \mu) } \\
 \chi_0({\bf r},{\bf r}' ; i\omega)  
    &= 2 \sum_{ja}   \frac{   {\left(\phi_j^{\mathrm{KS}}({\bf r})\right)}^*  \phi_a^{\mathrm{KS}}({\bf r})        {\left(\phi_a^{\mathrm{KS}}({\bf r}')\right)}^*  \phi_j^{\mathrm{KS}}({\bf r}')   }{ i\omega - ( \varepsilon_a^{\mathrm{KS}} - \varepsilon_i^{\mathrm{KS}}) } + c.c.
\end{align*}
where $G$ is the time-ordered Green's function and $\chi_0$ the independent-electron susceptibility used to build, within the random phase approximation, the screened Coulomb potential $W =  v + v \chi_0 W$ with $v$   the bare Coulomb potential. In the preceding equations,   $\eta$ is a positive infinitesimal and $\mu$ the chemical potential. Following quantum chemistry notations, (j/a) index occupied/empty energy levels. The $GW$ exchange correlation self-energy operator reads:
\begin{align*}
\Sigma^{GW}({\bf r}, {\bf r}' ; E ) = \frac{i}{2\pi} \int d\omega \; e^{i \eta \omega}
G({\bf r}, {\bf r}' ; E + \omega) W({\bf r}, {\bf r}' ;   \omega)
\end{align*}
where $W$ is obtained along the real axis by an analytic continuation (AC) from the $W(i\omega)$ calculated at $n_\omega$=14 optimized frequencies along the imaginary axis following a recently improved AC scheme  \cite{Fri19,Duc20,Duc21}. The knowledge of $\Sigma^{GW}$ allows to obtained $GW$ quasiparticle energies:
\begin{align*}
   \varepsilon_n^{GW} = \varepsilon_n^{\mathrm{KS}} + \left\langle \phi_n^{\mathrm{KS}} \left| \Sigma^{GW} \left( \varepsilon_n^{GW} \right) - V_{\mathrm{XC}}^{\mathrm{DFT}} \right| \phi_n^{\mathrm{KS}} \right\rangle
\end{align*}
We explicitly correct 50 occupied and 50 unoccupied states around the gap, states at lower/higher energy being corrected following the correction of the lowest/highest explicitly corrected level. 

Following extensive benchmarks on finite size systems \cite{Jac15a,Jac15b,Jac17}, we adopt for accuracy an eigenvalue self-consistent scheme (labeled ev$GW$) where the corrected quasiparticle energies are reinjected self-consistently in place of the input KS energy levels to calculate $G$ and $\chi_0$. Besides accuracy, such a self-consistent scheme allows to significantly reduce the impact of the arbitrariness associated with the choice of the DFT $V_{\mathrm{XC}}^{\mathrm{DFT}}$ functional used to generate the input $\lbrace \varepsilon_n^{\mathrm{KS}} , \phi_n^{\mathrm{KS}} \rbrace$ eigenstates. Following again recent benchmarks \cite{Jac15a,Jac15b,Jac17}, we select the global hybrid PBE0 functional for this initial step \cite{Ada99}. Finally, Coulomb integrals are expressed using   Coulomb-fitting resolution-of-the-identity techniques \cite{Whitten73,Ren_2012,Duc17} with standard auxiliary basis sets \cite{Weigend02} designed for the well-established cc-pVTZ \cite{Dunning89} and augmented \emph{aug}-cc-pVTZ basis sets  used below, or the auxiliary universal Weigend Coulomb fitting basis set \cite{Weigend08}  in conjunction with the smaller 6-311G(d) triple-zeta  plus polarization basis set \cite{Krishnan80} adopted for exploring large size systems. In the calculation of the susceptibility and self-energy, all unoccupied levels are considered.

 Once the ev$GW$ quasiparticle energies and screened Coulomb potential are obtained, singlet optical excitation energies 
 $\lbrace \Omega_{\lambda} \rbrace$ can be calculated within the Bethe-Salpeter equation (BSE) formalism that expresses the two-body electron-hole excitation eigenstates on the basis of the occupied-to-virtual level transitions:
\begin{align*}
\psi^{\mathrm{BSE}}_{\lambda}({\bf r}_\mathrm{e},{\bf r}_\mathrm{h}) = \sum_{ia} \left[ X_{ia}^{\lambda} \phi_i({\bf r}_\mathrm{h}) \phi_a({\bf r}_\mathrm{e}) +
 Y_{ia}^{\lambda} \phi_a({\bf r}_\mathrm{h}) \phi_i({\bf r}_\mathrm{e}) \right]
    \end{align*}
 where the $Y_{ia}^{\lambda}$ components indicate that we go beyond the Tamm-Dancoff approximation.
 These eigenstates solve a Casida-like \cite{Casida95} equation :
 \begin{equation} \label{eq:BSE-eigen}
    \begin{pmatrix}
		R & C  
		\\
		-C^*  & -R^{*}
	\end{pmatrix}
    \begin{pmatrix}
		X^{\lambda}
		\\
		Y^{\lambda}
	\end{pmatrix}
	=
	\Omega_{\lambda}
    \begin{pmatrix}
		X^{\lambda}  
		\\
		Y^{\lambda}
	\end{pmatrix},
\end{equation}
where the leading resonant block reads:
\begin{equation}
    R_{ia,jb} = \left( \varepsilon_a^{GW} - \varepsilon_i^{GW} \right) \delta_{ij} \delta_{ab} + \kappa V_{ia,jb}^\mathrm{x} - W_{ij,ab}^\mathrm{d}
    \label{eqn:resonant}
\end{equation}
with $\kappa$=2 or 0 for singlet or triplet excitations, and :
\begin{align*}
    V_{ia,jb}^\mathrm{x} = (ia|jb) = \int \mathrm d{\bf r} \mathrm d{\bf r}' \;
    \phi_i( {\bf r} ) \phi_a( {\bf r}  ) v({\bf r}, {\bf r}')
    \phi_i( {\bf r}' ) \phi_b( {\bf r}' ) \\
    W_{ij,ab}^\mathrm{d} = \int \mathrm d{\bf r} \mathrm d{\bf r}' \;
    \phi_i( {\bf r} ) \phi_j( {\bf r}  ) W({\bf r}, {\bf r}'; \omega = 0)
    \phi_a( {\bf r}' ) \phi_b( {\bf r}' )
\end{align*}
Good convergence can be obtained by including all occupied-to-virtual transitions such that $\left( \varepsilon_a^{GW} - \varepsilon_i^{GW} \right) \le E_{\mathrm{cut}}$ with $E_{\mathrm{cut}}$ = 30\,eV  (see SM~\cite{supplemental} for convergence study). Otherwise stated, this will be the cutoff adopted. Further, the convergence  with basis sets size is detailed in the SM~\cite{supplemental} evidencing the slower convergence of the BSE data as compared to TD-DFT. 

For computing the zero-point vibrational energy (ZPVE) associated with the ground and excited states, we take advantage of the analytic DFT and TD-DFT Hessian available in Gaussian 16.A.03 \cite{Gaussian16}. For these calculations, performed at the PBE0 \cite{Ada99} 6-311G(d) level in gas phase, we optimized the geometries with \emph{tight} convergence criteria for both the SCF and residual forces, and select the so-called \emph{ultra-fine} DFT integration grid. The second-order coupled-cluster, CC2 \cite{Koc95},  and algebraic diagrammatic construction,  ADC(2) \cite{Dre15}, calculations used for accuracy validation, are performed with the Turbomole 7.3 code \cite{Turbomole}, using various Dunning's basis sets and applying the RI approach with the corresponding auxiliary basis. Default Turbomole parameters and algorithms are applied. The calculations of the vibronic couplings and Huang-Rhys factors was achieved using the FCClasses-3 package \cite{San07a,San07b,San16b,Cer21}, on the basis of the vibrational data associated with the ground and excited states as obtained with the Gaussian code. These vibronic calculations, performed in a time-independent (TI) approach, use the Franck-Condon (FC) approximation and account for the so-called Duschinsky effects as well as temperature (298\,K), and are based on the most refined adiabatic Hessian (AH) model to determine the vibronic couplings  \cite{San16b}. We used 10$^8$ integrals in the TI scheme, allowing a good convergence (FC recovery $>$ 0.9). 

\section{Systems studied}

\begin{figure}[t]
	\includegraphics[width=8.6cm]{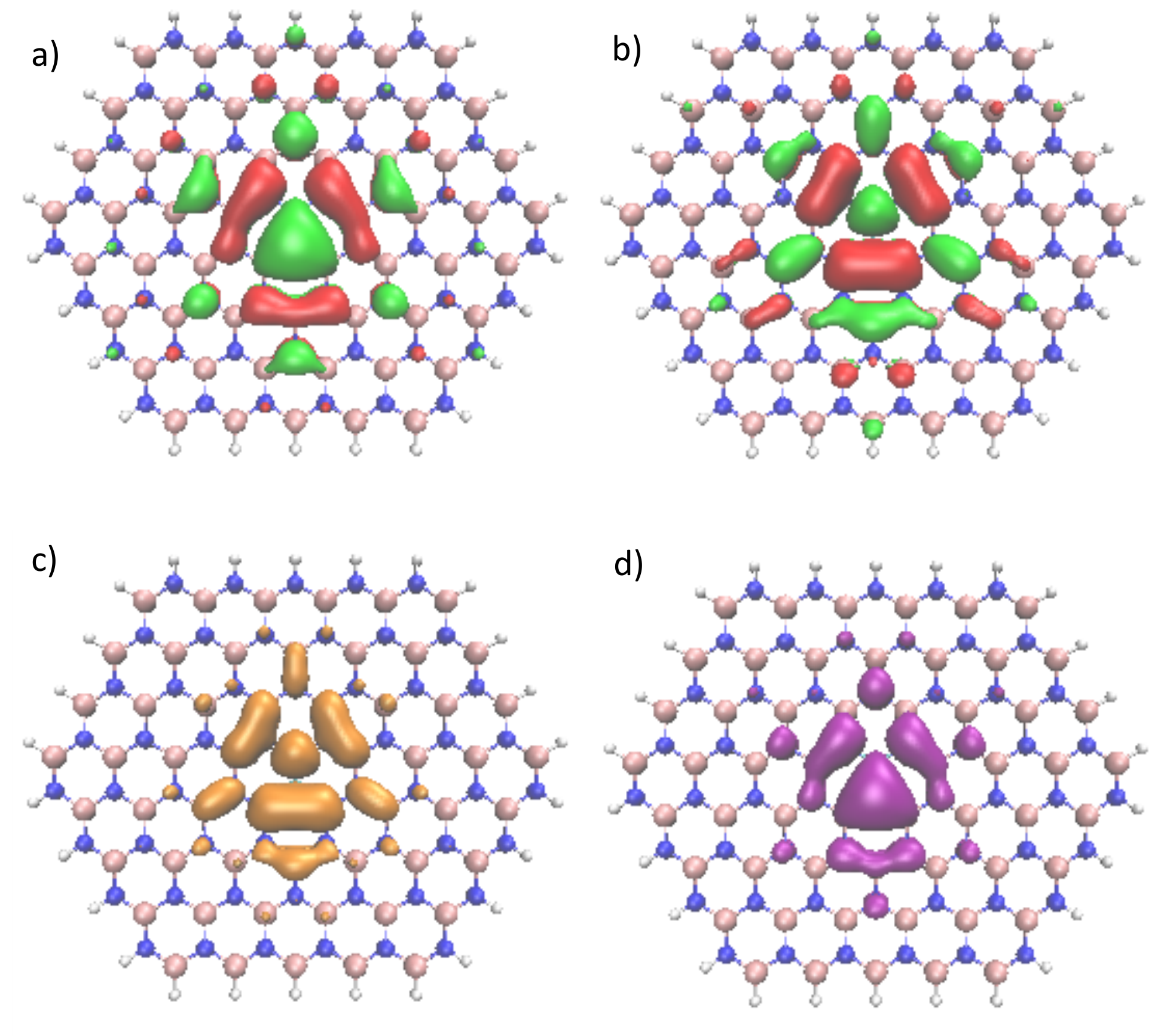}
	\caption{Isocontour representation for the 6-311G(d) PBE0 (a) HOMO and (b) LUMO levels in the C2@BN138 intermediate flake. In (c,d)  the  hole-averaged electron (orange) and electron-averaged hole (purple) densities, respectively, are given for the lowest singlet BSE  excitation.}
	\label{fig:HLplots}
\end{figure} 

The finite size \textit{h}-BN flakes with a central C-C dimer in substitution are represented in the Supplemental Material (SM) \cite{supplemental}. One of them, of intermediate size, is shown in Fig.~\ref{fig:HLplots}. The systems studied comprise from 42 to 240 B,C,N atoms, or 58 to 278 atoms including the H atoms used to avoid dangling bonds at the borders.  These systems are labeled C2@BNX where X counts the total number of atoms in the systems. The smallest flake is thus named C2@BN58 and the largest C2@BN278. In such systems, the minimal number of bonds between a central C atom and an edge B/N atom ranges from 4 to 11. 

The geometries of all systems were optimized at the PBE0/6-311G(d) level \cite{Ada99}. The corresponding C-C bond length is found to range between 1.368\,\AA{} and 1.369\,\AA{} in the smallest and largest flakes, respectively, in close agreement with the 1.361\,\AA{} periodic boundary condition PBEh(0.4)+D3  value of Ref.~\citenum{Mackoit_2019}, where PBEh(0.4) indicates a global hybrid in the PBE0 family with 40\,\% of exact exchange (instead of the default 25\,\%), while the D3 points to the Grimme empirical correction scheme for dispersion energies. 
For sake of illustration, we plot in Fig.~\ref{fig:HLplots}(a,b)  the PBE0 HOMO and LUMO levels associated with the intermediate size C2@BN138 system. 
As expected, independently of the system size, the PBE0 KS HOMO and LUMO levels are localized on the C-C carbon dimer. The bonding/anti-bonding character of the HOMO/LUMO along the C-C bond is clearly seen in Fig.~\ref{fig:HLplots}(a,b). 
 
Anticipating   on the analysis of the Bethe-Salpeter  optical excitation energies, we further plot in Fig.~\ref{fig:HLplots}(c,d) the hole-averaged $\rho_\mathrm{e}({\bf r})$ electron distribution and electron-averaged $\rho_\mathrm{h}({\bf r})$ hole distribution, with e.g.,
\begin{equation*}
    \rho_\mathrm{e}({\bf r}_\mathrm{e}) = \int \mathrm d{\bf r}_\mathrm{h} \; \left| \psi^\mathrm{BSE}_{\lambda=1}({\bf r}_\mathrm{e},{\bf r}_\mathrm{h} )\right|^2\,.
\end{equation*}
The electron and hole localization allows  to recognize that the lowest singlet BSE excitation is mainly of HOMO-LUMO character.  The weight $X_{ia}^{\lambda}$ squared of the BSE electron-hole wavefunction on the HOMO-LUMO transition is found to be consistently of the order of 0.94, rather independently of the system size beyond the C2@BN86 system, in close agreement with the 0.96 weight at the TD-PBE0 level. The remaining contributions, even though small, explain the evolution of the BSE excitation energy with respect to the $E_\mathrm{cut}$ parameter  (see SM \cite{supplemental}). While the two C atoms are non-equivalent, being bonded either to B or N atoms with different electronegativities, there is no sign of significant charge-transfer character for the HOMO-LUMO transition, as confirmed below by the large value of the oscillator strength. 

\section{Benchmark of methods}

We first benchmark the accuracy of the present BSE/ev$GW$ scheme, together with that of standard TD-DFT calculations using various exchange-correlation functionals, for the present carbon substituted \textit{h}-BN flakes. The study of finite size systems allows in particular to use accurate quantum chemistry wavefunction based methods, such as coupled-cluster (CC)   or algebraic diagrammatic techniques (ADC). A similar   comparison between mean-field DFT/TD-DFT approaches and wavefunction based methods on finite size systems  was already proposed in the case of a few defects in a   cluster geometry \cite{Reimers_2018,Korona_2019,Reimers_20}. Significantly more expensive than BSE and TD-DFT calculations, calculations with  the second-order wavefunction approaches, CC2 and ADC(2), with a large converged basis set (\emph{aug}-cc-pVTZ) are only possible for the smallest C2@BN58 system. Even though limited in size, this small system presents already well defined defect states, with a lowest singlet excitation energy within no more than 0.10--0.15\,eV of its limit value obtained in larger flakes (see below).


While very accurate ``theoretical best estimates" (TBE), such as those obtained from third order CC3 calculations \cite{Koc97,Ver21}, cannot be obtained for the present systems, CC2 and ADC(2) calculations have been shown, thanks to very large benchmarks on molecular families, to present mean absolute errors of the order of 0.10\,eV to  0.15\,eV as compared to TBE or experimental data for various type of excitations, from localized (Frenkel) to Rydberg or charge-transfer type transitions \cite{Win13,Jac15b,Oru16}. In particular, even though an approximation to CCSD,   CC2 provides a similar level of accuracy for localized (Frenkel) excitations  with a reduced cost scaling \cite{Koc97,Ver21}.  
 
Our data are presented in Table~\ref{tab:bench}. In the converged \emph{aug}-cc-pVTZ limit, we observe that the Bethe-Salpeter vertical excitation energy relying on self-consistent ev$GW$@PBE0 input quasiparticle energies and screened Coulomb potential, is in close agreement with the CC2 value, even though 0.1\,eV larger than its ADC(2) analogue. In other words, BSE falls within the error bar of second-order wavefunction schemes. 
The present results confirm  the accuracy of the  BSE/ev$GW$ approach, already demonstrated in large benchmarks on molecular systems \cite{Jac15a,Jac17} .


\begin{table}
	\caption{Lowest singlet ($S_1$) excitation energy for the C2@BN058 system. We compare quantum chemistry wavefunction methods [CC2 and ADC(2)], the many-body BSE/ev$GW$@PBE0 method, and TDDFT with the PBE, PBE0 and PBEh(0.4) functionals at the \emph{aug}-cc-pVTZ level.  Energies are in eV and oscillator strength in parenthesis.}
	\label{tab:bench}
        \begin{tabular}{c|c|c}
            CC2    &   ADC(2)      & BSE/evGW@PBE0         \\
            \hline
            4.761 (0.494) & 4.660 (0.468) & 4.765 (0.407)         \\
            \multicolumn{3}{c}{   }                                         \\
            %
            %
            PBE0         &  PBE          & PBEh(0.4)             \\ 
            \hline
            4.716 (0.406) & 4.024 (0.119) & 4.966 (0.443)
    	\end{tabular}
\end{table}


Among the TD-DFT calculations, we further observe that the standard TD-PBE0, with its 25\,\% ratio of exact exchange, offers an excellent accuracy, standing in between ADC(2) and CC2 data. TD-PBE0 and BSE/ev$GW$ excitation energies fall within 50\,meV of each other, while their oscillator strengths are also in excellent agreement. Such values are also in good agreement with the 4.78 eV CAM-B3LYP energy of Ref.~\cite{Korona_2019} obtained with a minimally augmented cc-pVDZ basis for a cluster equivalent in size to our C2@BN138 system. In contrast, TD-PBE calculations yield an excitation energy that is significantly too small, together with an oscillator strength that is a factor 4 smaller than that of the CC2 or ADC(2) references. On the contrary, adopting the PBEh(0.4) functional with 40\,\% of exact exchange  yields an excitation energy that is somehow too large. The balanced performance of the PBE0 functional is consistent again with large TD-DFT benchmark on molecular systems, proving that PBE0 is one of the most accurate functional for localized excited states \cite{Lau13}. We adopt thus the  BSE/ev$GW$ and TD-PBE0 schemes for exploring now the convergence with system size. Further, since analytic forces in the excited states have not been developed at the BSE level, reorganization energy  and vibrational-spectrum calculations in the excited state are conducted below at the TD-PBE0 level. 

\section{Vertical excitation energies}

We now explore the evolution of the lowest singlet ($S_1$) vertical excitation energy with respect to system size. While TD-DFT and BSE data for the smallest C2@BN58 system have been explicitly calculated at the \emph{aug}-cc-pVTZ level, calculations on the largest flakes have been performed with the  6-311G(d) Pople atomic basis set level, stopping at the C2@BN202 system for BSE calculations. As shown in the SM \cite{supplemental}, the evolution with system size of the excitation energy is weakly dependent on the basis size, allowing to safely extrapolate the 6-311G(d) data to their \emph{aug}-cc-pVTZ value. Further, the small evolution of these energies beyond the C2@BN86 system insures that errors in the extrapolation are very limited. 

\begin{figure}[t]
	\includegraphics{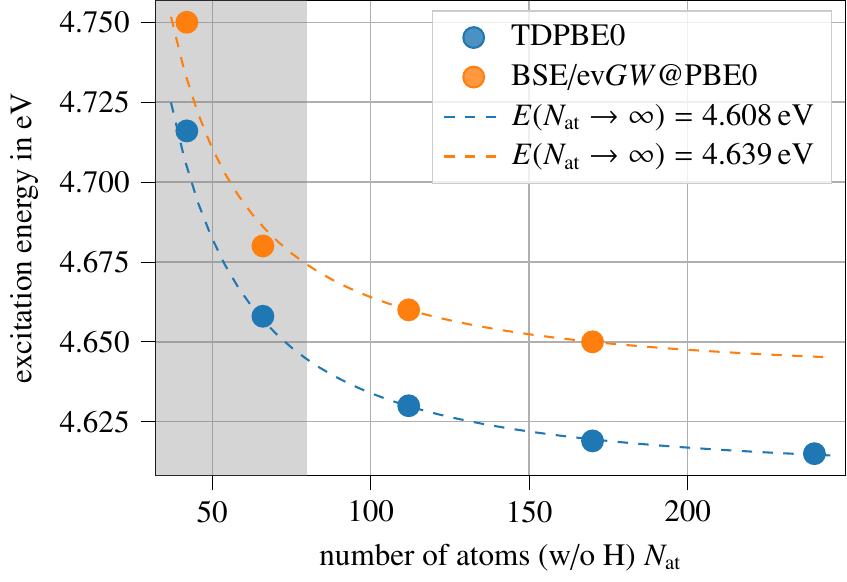} 
	\caption{Evolution of the lowest vertical singlet ($S_1$) excitation energy (in eV) with respect to the number of non-H atoms ($N_{\mathrm{at}}$). TD-PBE0 (orange) and BSE (blue) data are presented. Plotted values have been extrapolated to the \emph{aug}-cc-pVTZ level (see text and the SM \cite{supplemental}). The largest flakes data points (out of the shaded area)  are fitted by an asymptotic  $\; f(x)=a \cdot x^3 + c \; $ functional with $x = 1/ \sqrt{N_{\mathrm{at}}}$  where $N_{\mathrm{at}} \sim  {R^2}$   with $R$ the flake average radius. The infinite radius extrapolated energy value is indicated in the Inset.}
	\label{fig:eS1vsize}
\end{figure} 

As shown in Fig.~\ref{fig:eS1vsize}, both the TD-PBE0 and BSE vertical excitation energies decrease  with system size, converging relatively fast,  with the C2@BN138 system being within $\sim$20\,meV of the extrapolated infinite size limit.  The decay rate within the TD-PBE0 and BSE formalisms are  similar.    
Several effects can be invoked to understand the reduction of the excitation energy with system size. A first possible explanation could hinge on the confinement of the defect states, resulting in an increase of the HOMO-LUMO gap in the small size limit. The analysis of the KS HOMO and LUMO levels  indicates a 75\,meV reduction of the gap between the C2@BN58 and C2@BN278 systems. Even though not directly related to the optical gap, this reduction of the Kohn-Sham gap, that fully accounts for confinement effects, can contribute to the $\sim$0.11\,eV decrease of the optical gap with system size. Confinement is expected to yield an evolution of the band gap that scales as $1/R^2$ where $R$ is the flake radius.

\begin{figure}[t]
	\includegraphics{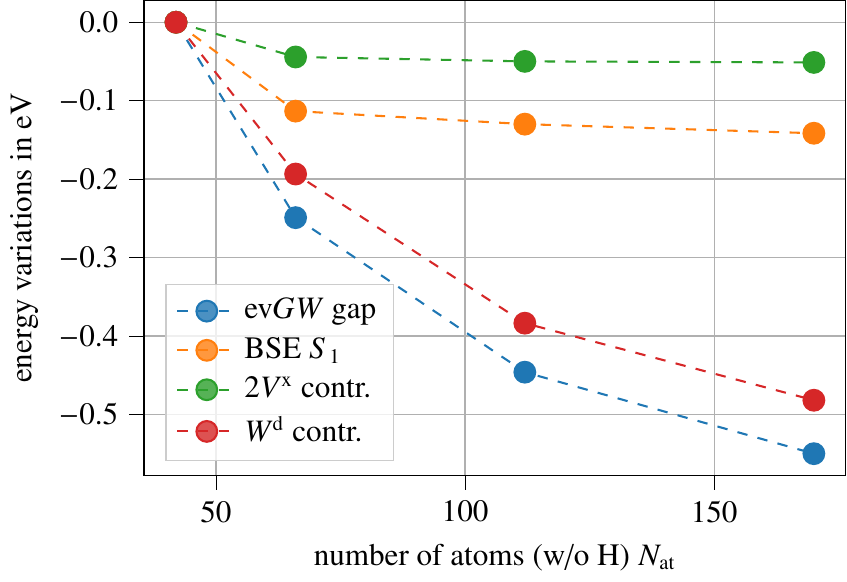} 
	\caption{Size evolution  with respect to the smallest C2@BN58 system of the ev$GW$ HOMO-LUMO gap (blue),  the lowest vertical singlet ($S_1$) excitation energy (orange), the $\langle \psi^{\mathrm{BSE}}_1 | 2V^\mathrm{x} |\psi^{\mathrm{BSE}}_1 \rangle$ (green) and $\langle \psi^{\mathrm{BSE}}_1 | W^\mathrm{d} |\psi^{\mathrm{BSE}}_1 \rangle$ (red) matrix elements. Calculations performed with the 6-311G(d) atomic basis set. Energies are in eV.}
	\label{fig:variations}
\end{figure} 

Another source of reduction of the optical gap with system size is the polarization energy, or the increase of environmental screening upon increase of the system size.  
It is interesting to analyze in this respect the evolution with size of the various contributions to the BSE absorption energy, including in particular the ev$GW$ HOMO-LUMO gap  together with the expectation values $\langle \psi^{\mathrm{BSE}}_1 | 2V^\mathrm{x} |\psi^{\mathrm{BSE}}_1 \rangle$  and  $\langle \psi^{\mathrm{BSE}}_1 | W^\mathrm{d} |\psi^{\mathrm{BSE}}_1 \rangle$   of the bare and screened Coulomb potentials over the $S_1$ BSE eigenstate (see Eqn.~\ref{eqn:resonant}). As shown in Fig.~\ref{fig:variations}, the drop of the ev$GW$ gap is much larger than the 75\,meV calculated at the KS level. The $GW$ formalism properly mimics a photoemission experiment that measures charging energies. Upon charging the defect levels, the generated Coulomb field strongly polarizes the surrounding medium. The environment charge rearrangement generates a reaction field that stabilizes the added charge. This effect, that closes the gap, is a strong long range polarization effect that slowly converges with system size.  

The much faster convergence of the BSE excitation energy can be explained by the evolution of the screened electron-hole $\langle \psi^{\mathrm{BSE}}_1 | W^\mathrm{d} |\psi^{\mathrm{BSE}}_1 \rangle$ term that decays with system size similarly as the ev$GW$ energy levels, compensating largely the drop in the gap. In other words, the evolution of the screened Coulomb potential with respect to the size of the polarizable environment affects in a similar fashion both the   electron-electron (exchange-correlation) interaction  that enters the ev$GW$ gap, and the electron-hole interaction that is subtracted in the BSE Hamiltonian. The unscreened bare electron $\langle \psi^{\mathrm{BSE}}_1 | 2V^\mathrm{x} |\psi^{\mathrm{BSE}}_1 \rangle$ contribution varies much less dramatically with system size, confirming that it is the change in long-range screening that determines the evolution of the ev$GW$ gap and the counterbalancing electron-hole interaction. Overall, the optical excitation is a neutral excitation that polarizes the environment much more weakly as compared to charged excitations.

To analyze more precisely the scaling properties of the optical excitation energy, we briefly recall the so-called linear-response (LR) and state-specific (SS) contributions to the energy shift induced by a polarizable environment on a localized excitation \cite{Cammi05,Duc18,Gui19}.
The localized excitation is here the dimer-defect excitation and the polarizable environment is the \textit{h}-BN host. 
Assuming for simplicity a pure HOMO-LUMO transition, the LR contribution to the shift of the absorption energy induced by the environment reads $ \langle \rho_{\mathrm{HL}} | v^{\mathrm{reac}} | \rho_{\mathrm{HL}} \rangle $, with $ \left| \rho_{\mathrm{HL}}\right\rangle = \left|\phi_\mathrm{H} \phi_\mathrm{L}\right\rangle $ the product of the HOMO and LUMO one-body orbitals. The reaction field $ v^{\mathrm{reac}}({\bf r},{\bf r}')$ describes the potential generated in ${\bf r}'$ by the variation of charge induced in the environment by a  point charge added in ${\bf r}$. In the LR contribution, the environment reacts to the $\rho_{HL}$ transition codensity, and the LR effect is closely related to the magnitude of the oscillator strength. In the SS contribution, the energy shift reads $\langle \Delta \rho_{\mathrm{ES}} | v^{\mathrm{reac}} | \Delta\rho_{\mathrm{ES}} \rangle / 2$, with $ \Delta\rho_{\mathrm{ES}}$ the variation of the total charge density between the ground-state and the excited state, i.e., the SS contribution is large when the density reorganization is large, such as in CT excitations. Both contributions scale as the reaction field that we now show to vary  as $1/R^{3}$ with $R$ the radius of the flake as expected in 2D systems.  

Both the $\rho_{\mathrm{HL}}$ and $\Delta \rho_{\mathrm{ES}}$ density variations have no monopolar component. However, the rather large oscillator strength associated with the transition indicates a large $\rho_{\mathrm{HL}}$ dipolar component.  The   $1/r^3$ electric field associated with these dipolar  components induces dipoles proportional to $\alpha_{\mathrm{at}} / r^3$ on B/N atoms located at a distance $r$ of the defect, where $\alpha_{\mathrm{at}}$ is some effective average atomic polarizability. Each induced dipole generates a reaction  potential that stabilizes the defect excited-state dipole. Integrating over the reaction field generated by all B/N atoms in the flake leads indeed to an overall reaction potential that  scales as $1/R^{3}$, with  $R \simeq \sqrt{N_{\mathrm{at}}}$, taking $N_{\mathrm{at}}$ to be the number of B/C/N atoms. Fitting with such a functional form the  TD-PBE0 data for the 3 largest flakes   (Fig.~\ref{fig:eS1vsize}, dashed blue line) leads to a fit that smoothly goes close to the 2 smallest systems for which such an asymptotic behavior may be less  accurate, accounting in particular for the $1/R^2$ behavior associated with confinement. Fitting the 2 largest systems for the BSE data leads to a fit somehow less satisfactory in the small size limit (Fig.~\ref{fig:eS1vsize}, dashed orange line). We must note that such an asymptotic behavior relies on the assumption that the defect excitation $\rho_{\mathrm{HL}}$ and $ \Delta\rho_{\mathrm{ES}}$ charge variations do not change with system size, a condition that may not be fully validated in the small size limit. 

Overall, the large system size extrapolated TD-PBE0 vertical $S_1$ excitation energy is found to be 4.61\,eV, within 5\,meV of the value obtained for the largest C2@BN278 system. Similarly, the infinite-system size extrapolated BSE value of 4.64\,eV lies within 10\,mev of the value obtained for the C2@BN202 system. We adopt in the following these extrapolated values. The present results confirm the conclusions of Ref.~\citenum{Reimers_20} that optical excitations converge rather fast with cluster size. 


\section{Reorganization energies and ZPVE corrections}

\begin{figure}[t]
	\includegraphics{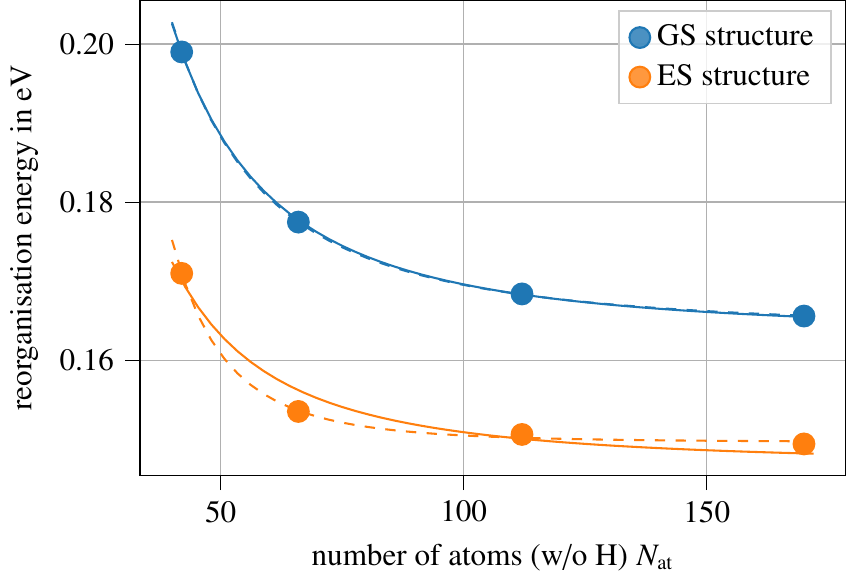} 
	\caption{Evolution of the ground and excited state reorganization energy (in eV)  at the TD-PBE0/6-311G(d) level.
	The solid lines are a 1/$N_{\mathrm{at}}^2$ fit to the data points. The dashed lines are a  "free" exponent 1/$N_{\mathrm{at}}^{\alpha}$ with 
	$\alpha \simeq 2.08$ for the ground-state and $\alpha \simeq 3.71$ for the excited-state. Free fit infinite size extrapolated values amount to 0.164\,eV for the ground-state and 0.150\,eV for the excited-state, respectively. 
	\label{fig:FC}} 
\end{figure}

As a next step, we explore the convergence with system size of the  reorganization energies  in the ground  and excited states, namely the differences $\;  \lambda^{GS} = E_{\mathrm{GS}}[R_{\mathrm{ES}}] - E_{\mathrm{GS}}[R_{\mathrm{GS}}] \;$
and $\; \lambda^{ES} = E_{\mathrm{ES}}[R_{\mathrm{GS}}] - E_{\mathrm{ES}}[R_{\mathrm{ES}}] \;$ where $R_{\mathrm{GS}}$ and $R_{\mathrm{ES}}$ indicate the $S_0$ and $S_1$ geometries, respectively. Analytic forces in the ground and excited states are not available within the BSE formalism so that we use TD-PBE0 for this exploration. As shown in Fig.~\ref{fig:FC}, the reorganization energies slightly decrease with system size  with variations that are of the same magnitude as those evidenced above for the vertical excitation energy. 
While an elastic theory analysis of the deformation energy induced by a point defect in 2D finite-size systems is beyond the present study, we find that 
the ground-state reorganization energy can be  accurately fitted by a $1/N_{\mathrm{at}}^{2}$ or $1/R$ law, where $N_{\mathrm{at}}$ in the number of atoms and $R$ the effective radius of the flake. 
 We extrapolate the ground-state reorganization energy to 0.164\,eV. For the ES, the reorganization energy is found to be somehow smaller 
 with a decay law that seems to rely on a larger exponent. 
Whatever the reliability of the fit, the extrapolated value and the one explicitly calculated for the largest C2@BN202 system are within a very few meV from each other, indicating again that any extrapolation inaccuracy can hardly affect the conclusions. 
We adopt the values of 0.16\,eV and 0.15\,eV for the reorganization energy in the ground and excited state, respectively. The smaller value of the reorganization energy in the excited state can be associated with smaller vibrational energies as compared to the ground state. This is what we now explore by considering the zero-point vibrational energy (ZPVE).

To further account for the ZPVE correction, we perform calculations of the vibrational spectrum in the ground-state at the ground-state geometry, and in the $S_1$ excited state at the excited-state geometry. The difference in ZPVE between the excited and ground states amounts to 0.147\,eV for the smallest C2@BN58 system and 0.131\,eV in the larger C2@BN86 system. As expected \cite{Goe09,Jac12d}, the vibrational modes in the excited states are ``softer'', consistent with the antibonding character of the LUMO level (see Fig.~\ref{fig:HLplots}b).
Such vibrational energy differences are rather large yet not exceptional \cite{Goe09,Jac12d}. This is likely related to  a significant tensile stress for the C-C bond embedded in the \textit{h}-BN lattice with larger bond length, a constraint that is essentially released upon populating an antibonding state. In fact, the C-C bond length changes from 1.37\,\AA{} to 1.45\,\AA{} at the $S_1$ excited state geometry $R_\mathrm{ES}$, a bond length that can be compared to 1.44\,\AA{}, the PBE0/6-311G(d) B-N bond length at the center of an undoped pristine \textit{h}-BN flake. In the excited state, the C-C bond length is thus much closer to that of the host network, confirming the release of the tensile stress.  
Assuming that the ZPVE correction will further decrease slightly in the limit of larger systems, we adopt in the following the value of 0.13\,eV. 

\begin{figure}
    \centering
    \includegraphics[width=8.6cm]{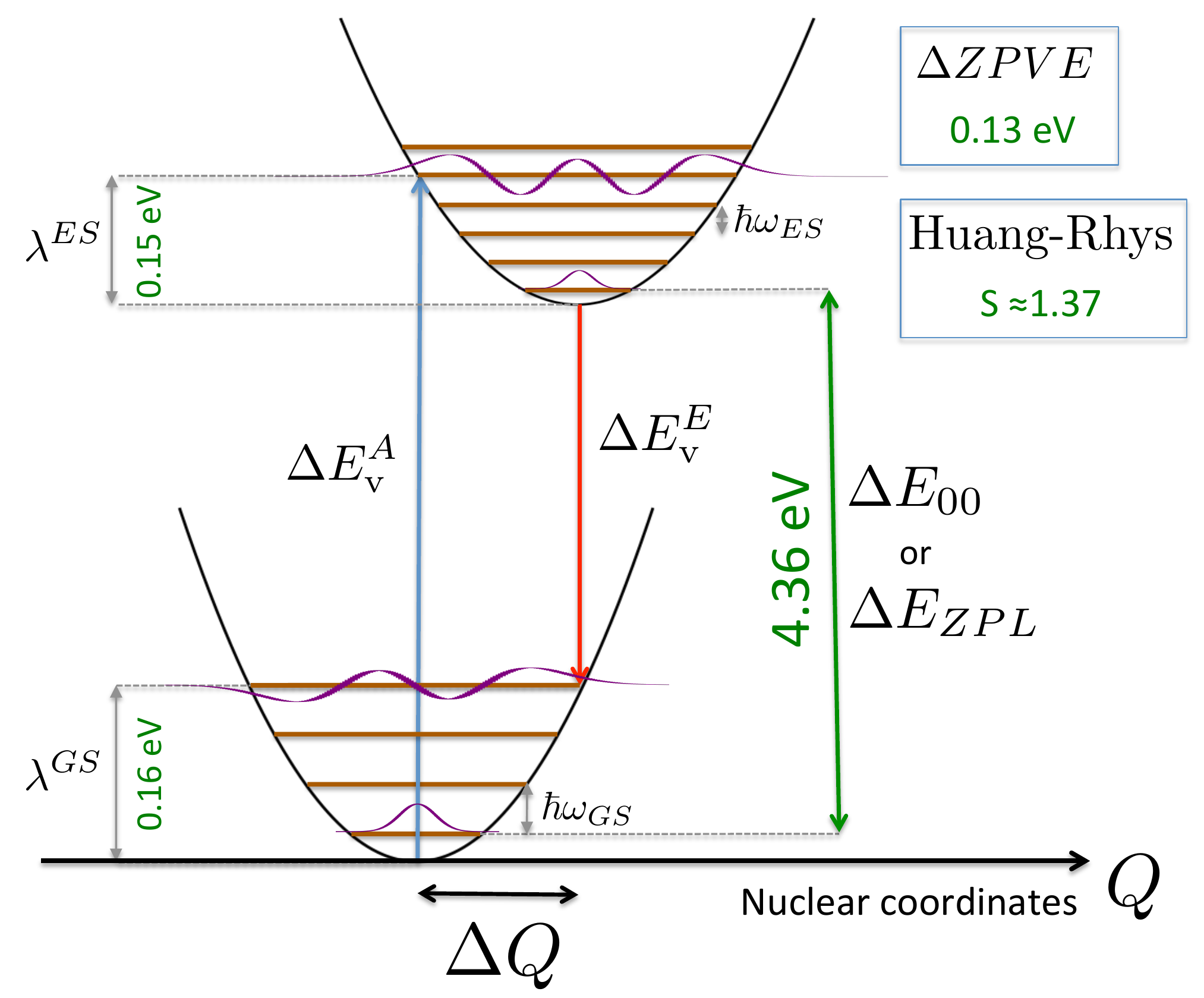}
    \caption{
    Symbolic representation of the lowest singlet vertical and zero-phonon transitions, with the reorganization energy in the ground  ($\lambda^\mathrm{GS}$) and excited  ($\lambda^\mathrm{ES}$) states, ZPVE contribution and Huang-Rhys factor $S$.
    }
    \label{fig:sketch_FC+ZPVE}
\end{figure}

Finally, taking the 4.64\,eV calculated BSE vertical absorption energy, subtracting the 0.15\,eV reorganization energy in the $S_1$ excited state and the ZPVE difference contribution of 0.13\,eV, we obtain a zero-phonon-line (ZPL) 
energy of about 4.36\,eV (see summary in Fig.~\ref{fig:sketch_FC+ZPVE}). This is close to the 4.31\,eV value found in Ref.~\citenum{Mackoit_2019} using constrained DFT calculations in a PBC geometry, even though this previous value does not contain the ZPVE contribution.

\section{Huang-Rhys and Debye-Waller factors  }

We now discuss the Huang-Rhys and Debye-Waller factors as an additional mean to connect to experimental data~\cite{Vokhmintsev_2019}.
Closely related to the reorganization energy from the ground-state to the excited-state geometry, the Huang-Rhys (HR) factor is a measure of the strength of coupling between the excitation and the vibrational modes. In our case, we computed the HR factors for all vibrational modes using the adiabatic Hessian vibronic model relying on the (TD-)PBE0/6-311G(d) vibrational data of the two states.
The HR factor reported below, $S$, is obtained as the sum of the HR terms determined for all vibrational modes.

The $S$ factor associated with the smallest C2@BN58 system is found to be $S=1.37$, decreasing to $S=1.23$ in the case of the C2@BN86 system. This 10\,\% decrease with system size is consistent with the 10\,\% decrease of the reorganization energy between these two systems in the excited state (see Fig.~\ref{fig:FC}). The very limited decrease of the reorganization energy at larger sizes indicates that the value obtained for the C2@BN86 system is a very good estimate of the converged $S$ in the large size limit. Our calculated value falls within the 1--2 range of the   experimental estimate \cite{Vokhmintsev_2019}. The related $W_{ZPL} \approx e^{-S}$ Debye-Waller coefficient is found to be 0.29, in good agreement with  the estimated 0.26 experimental value \cite{Vuong16}.

In the C2@BN58 system, the largest single mode contribution to the total $S$ factor is a clear C-C stretching mode located at 1347\,cm$^{-1}$ (168\,meV) in the excited state, that according to the Duschinsky analysis, couples to several ground state mode, the strongest being the alike C-C stretching mode at 1644\,cm$^{-1}$ (204\,meV)   that  was shown in Ref.~\cite{Korona_2019} to stand higher in energy as compared to the optical modes of pristine \textit{h}-BN. The second and third contributions originate in couplings between low-frequency breathing modes: the excited state modes appearing at 338\,cm$^{-1}$ (42\,meV)  and 236\,cm$^{-1}$ (29\,meV) being respectively coupled to the ground state modes at 254\,cm$^{-1}$ (31\,meV)  and 242\,cm$^{-1}$ (39\,meV). In the C2@BN86 flake, the largest single mode contributions to $S$ are coming from, on the one hand, a breathing mode (272\,cm$^{-1}$ or 33\,meV in the excited state), and on the other hand, the excited-state C-C stretching at  1260\,cm$^{-1}$ (156\,meV). These modes are represented in the SM~\cite{supplemental}. 
In the computed vibrationally-resolved emission spectrum (see the SM~\cite{supplemental}), the strongest ``stick'' contributions also involve these vibrations. We note that the energy of the stretching and breathing modes can be associated with the 174\,meV and 60\,meV  energy spacing between phonon replica of the 4.08\,eV zero-phonon line as observed in the experimental PL spectrum of Ref.~\citenum{Vokhmintsev_2019}. 

\section{Multi-layer polarization effects}
\label{section:multilayer}

\begin{figure}[t]
	\includegraphics[width=8.6cm]{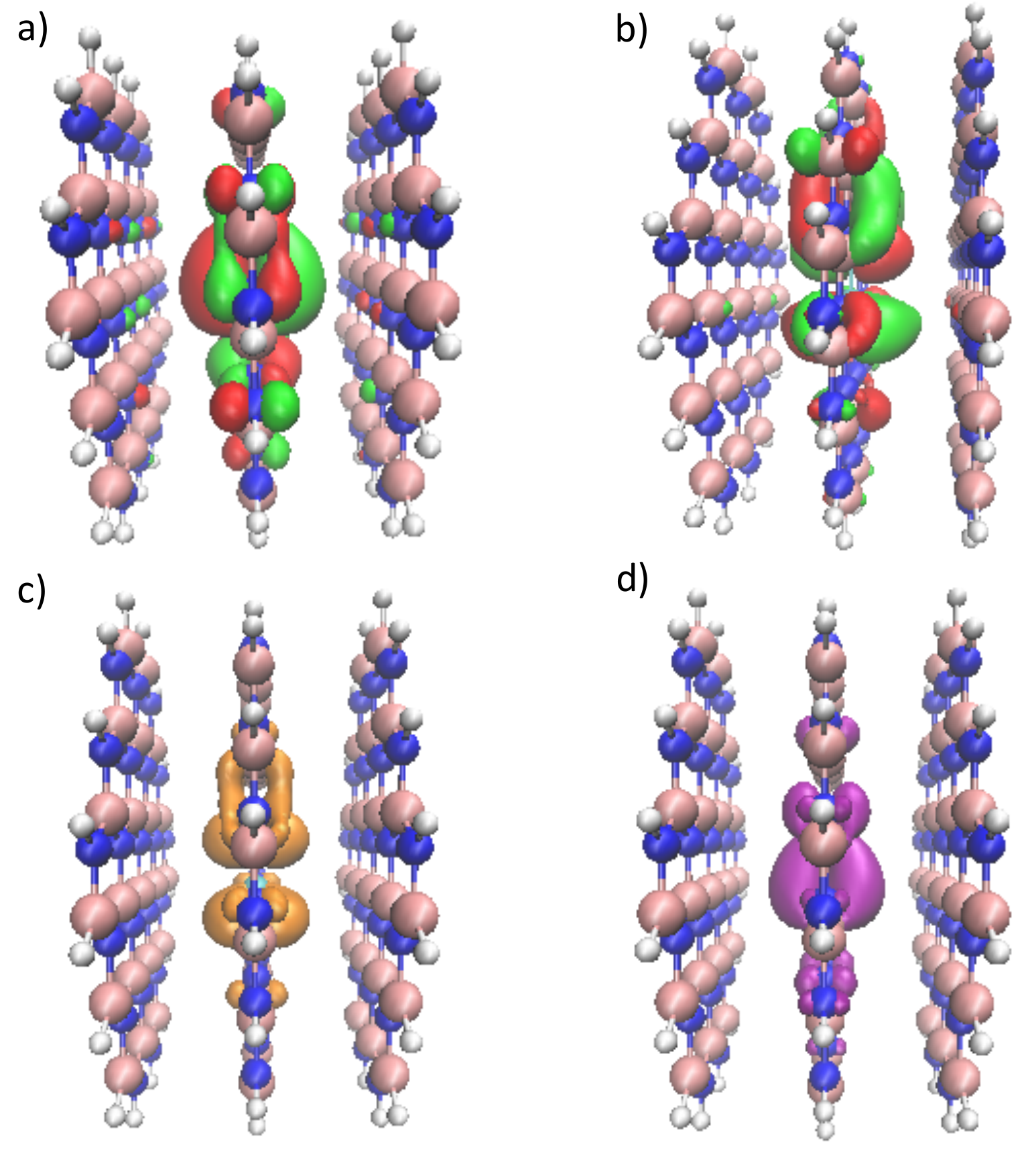}
	\caption{Isocontour representation of the defect-associated (a) HOMO and (b) LUMO levels in a C2@BN58 trilayer system with the defect localized in the inner layer. The stacking is of AA' type and the interlayer spacing has been set to 3.33\,\AA{}.  In c) and d), the hole-averaged electron density and electron-averaged hole density for the lowest defect-associated BSE singlet excitation  are represented.}
	\label{fig:trilayer}
\end{figure}

The theoretical $\sim$4.36 eV zero-phonon emission energy calculated for the defected monolayer system remains larger than the   4.08\,eV experimental value. However,
experimental data  were not  acquired for defected single-layer \textit{h}-BN sheet but multilayer systems.
As such, environmental screening phenomena are expected to affect the optical transition energies, an effect that may be labeled a solvatochromic shift by analogy with the change in absorption or emission energies for solvated species as compared to the gas phase. 
 Such environmental screening phenomena were already shown in Refs.~\cite{Smart18,Wang19} to stabilize the defect states, namely reducing the   HOMO-LUMO photoemission gap. To study this effect in the present case of photoluminescence,  we explore again the optical properties of  our smallest C2@BN58 system in a bilayer and trilayer geometry.
For these model systems, the layers are individually relaxed  and then stacked without further relaxation.
We adopt the standard AA' geometry (B atoms on top of N atoms) with a 3.33\,\AA{} interlayer spacing \cite{Lin2010,Shi2010}. In this geometry, the C-C dimer is located on top of a B-N dimer of neighboring layers. As shown in Fig~\ref{fig:trilayer}, the defect-associated HOMO and LUMO levels remain strongly localized on the defected layer. 

The most salient feature is that the defect-associated BSE absorption energy is red-shifted by 0.17\,eV and 0.27\,eV from the monolayer to the bi- and trilayer systems, respectively, taking the central layer in its ground-state geometry. The very same 0.27\,eV redshift is found for the $S_1$ energy of a trilayer system taking the central layer in its $S_1$ excited state geometry. As such, the effect on the excitation energy of the neighboring layers seems rather independent of the central layer geometry, suggesting that the reorganization energy and ZPVE contributions calculated for the defected monolayer remain good estimates in the multilayer case.

To better analyze this evolution, we selected the central monolayer in its ground-state geometry and found that between the monolayer and the trilayer systems, the KS gap closes by  0.17\,eV while the $GW$ gap closes by 0.78\,eV.  As such, direct hybridization between layers is not negligible, but is marginal as compared to polarization effects, namely long-range screening by adjacent layers. 
 This confirms the stabilization of the defect states, namely the reduction of the ionization potential and increase of the electronic affinity, induced by environmental screening. The same effect was studied in the case of \textit{h}-BN layers on a substrate \cite{Wang19}.  
Again, interlayer screening reduces as well the electron-hole interaction, leading to a reduction of the optical excitation that is significantly smaller than that of the (photoemission) gap.  This remains however a non-negligible effect when attempting to corroborate theoretical with experimental data, a message that applies to the present carbon-dimer defect but certainly to any other defects. 

We can speculate that most of the screening effects, inducing a redshift of the optical lines as compared to the defected monolayer system, originates from the screening by a very few neighbouring layers, tentatively explaining why experiments performed on bulk or few-layers \textit{h}-BN provide a very stable ZPL emission line at about 4.1\,eV. 
Studying the evolution of the emission line with respect to the number of layers would be prohibitively expensive and lies beyond the scope of the present study. The main message however is that calculations of absorption or luminescence energies on monolayer systems should overestimate the experimental values  by a few tenths of an eV.  We adopt the plausible value of 0.3\,eV, namely slightly larger than the 0.27\,eV value explicitly calculated for the trilayer system. 

Reducing the calculated 4.36\,eV ZPL energy for the monolayer system by a redshift of the order of 0.3\,eV brings that calculated value to close agreement with the 4.08\,eV experimental line. Such a close agreement is certainly fortuitous given the $\simeq$0.1\,eV difference on the vertical absorption energies between high-level methods [BSE, CC2, ADC(2)], the rough estimate of the neighboring layers screening effect (rounded to 0.3\,eV) and the extrapolations needed to reach the infinite system size  and complete basis set limits. However, the present estimate certainly comforts the carbon-dimer defect as the origin of this emission line as proposed by the authors of Ref.~\citenum{Mackoit_2019}. 


\section{Photoemission gap and excitonic binding energy}  

We close this study by a brief discussion of the dopant-associated 
HOMO-LUMO gap as obtained within the ev$GW$ formalism and the resulting excitonic binding energy obtained from subtracting the optical gap. Since no experimental data are available for these quantities, the resulting values will stand as predictions. More details are available in the SM~\cite{supplemental}. 

Starting with the smallest C2@BN58 system, our most converged \emph{aug}-cc-pVTZ ev$GW$ value yields 8.28\,eV which is very close to the extrapolated  8.3\,eV to 8.4\,eV values in the complete basis set (CBS) limit.  
Adapting the discussion provided above for the scaling laws associated with long-range polarization effects, but treating here the response to  charged (photoemission) excitations,  the evolution with size of the gap scales as 1/$R^2$ (see e.g. Ref.~\citenum{D_Avino_2016} in the case of molecular 2D systems).  Fitting our $GW$ data points with such a functional form, the extrapolated gap in the infinite system size limit is found to close by 0.77\,eV as compared to the smallest C2@BN58 system,   leading to  $\simeq$7.6\,eV for the carbon dimer defect HOMO-LUMO gap in an infinite \textit{h}-BN monolayer.  Subtracting the 4.64\,eV vertical optical absorption energy, one obtains a very large $\simeq$3\,eV electron-hole binding energy. This is much larger than the estimated 0.7-0.8\,eV \cite{Paleari_2018,Aggoune_2018,Kolos_2019} and 1.9\,eV - 2.1\,eV \cite{Wirtz06,Galvani_2016,Cudazzo16,Paleari_2018} excitonic binding energy for layered and single-sheet undefected \textit{h}-BN systems. The clear localization of the Frenkel electron-hole pair on the defect, as shown in   Fig.~\ref{fig:HLplots}(c,d), can certainly explain such a large increase of the excitonic binding energy.  


 Finally, the knowledge of the defect $GW$ ionization potential and electronic affinity allows  estimating the defect charge
transition levels (CTLs) as an indication of the chemical potential energy range where the neutral defect is more stable than the charged ones (see e.g. Refs.~\cite{Wu_2017,Wang20}). Exploiting the fast convergence with system size of structural reorganization energies, we estimate from the large C2@BN138 system the reorganization energy from the neutral  to the cationic state geometry   to be 0.35\,eV [PBE0 6-311G(d) value]. Further, in the infinite size limit  (see SM~\cite{supplemental}, Section VIII), we find the position of the occupied defect level to be $\simeq$0.64\,eV above the monolayer \textit{h}-BN occupied band edge. These data bring  the (+1/0) CTL at about 1.0\,eV above the \textit{h}-BN top of the valence band, in close agreement with the DFT calculations of Ref.~\cite{Mackoit_2019}. Further, with a reorganization energy to the anionic geometry of 0.25\,eV and a defect HOMO-LUMO gap of 7.6\,eV, we find the (0/-1) CTL to lie at about 8.0\,eV above the valence band edge. These calculations confirm the outcomes of Ref.~\cite{Mackoit_2019} that for the isolated defected monolayer, the neutral state should be the most stable for a chemical potential located throughout most of the gap. We will not explore here the effect of multilayer screening on the CTL positions \cite{Wang20}.

\section{Conclusion}

We have studied the optoelectronic properties of the recently proposed carbon-dimer defect in hexagonal boron-nitride using the $GW$ and Bethe-Salpeter equation many-body Green's function formalisms in a finite size cluster (flake) approach.  More precisely, we calculate vertical excitation energies within the BSE formalism, while relaxation in the excited state and zero-point vibrational energies (ZPVE) are calculated at the TD-PBE0 level. Comparison with wavefunction CC2 and ADC(2) calculations on a small C2@BN58 atoms system validated the accuracy of both the BSE formalism and of the TD-PBE0 approach in the family of TD-DFT techniques. 

The study of all relevant quantities with respect to system size allows to find that the vertical optical absorption energy, the reorganization energy, and the ZPVE contributions, converge rather quickly with system size, being close to their extrapolated value for systems containing between 100 and 200 B/C/N atoms. As a result, we find a BSE ZPL emission energy of 4.36\,eV for the defected monolayer, accounting for a 0.15\,eV reorganization energy in the excited state and a significant 0.13\,eV ZPVE reduction from  the ground to the excited state. 
  We show  further that additional screening associated with neighboring layers reduce the defect emission energy, with a 0.27\,eV redshift in the simple case of a model trilayer system, bringing the calculated ZPL energy in close agreement with the 4.08 eV experimental value. Further, the S$\simeq$1.37 Huang-Rhys and $W_{ZPL} \simeq$0.25   Debye-Waller coefficients   are found to be in excellent agreement with experimental estimates.  The present results thus confirm the identification proposed in Ref.~\citenum{Mackoit_2019} that the carbon-dimer defect is at the origin of the 4.08\,eV ZPL luminescence line. On a more general ground, the present study suggests that    many-body Bethe-Salpeter calculations, performed on finite size systems containing a unique defect, stand  as a valuable alternative to periodic boundary approaches for the study of the optical properties of defects in the dilute limit.  

In contrast to the neutral optical excitations, the $GW$ quasiparticle energies (HOMO, LUMO, etc.), namely charged excitations, converge very slowly with system size.   Extrapolation of quasiparticle energies to the infinite size limit can however be safely obtained with a 1/$R^2$ law, with $R$ the radius of the 2D flake  containing the defect. This stems from a simple polarization scaling law in insulating 2D systems. As a result, the gap is found to close by $\simeq$0.77\,eV from the smallest C2@BN58 system to the infinite   size limit, leading to an ev$GW$  HOMO-LUMO gap of 7.6\,eV in the monolayer complete-basis set and infinite system size limits. Subtracting the vertical absorption energy, one obtains a very large excitonic binding energy of 3\,eV. The very localized Frenkel  character of the excitation can justify that this electron-hole interaction is much larger than what is found in pristine \textit{h}-BN systems where the excitons are much more delocalized.

\begin{acknowledgments}
 X.B., I.D., and D.J. thank the ANR for financial support in the framework of the BSE-forces grant (ANR-20-CE29-0005). Calculations have been performed on the French TGCC Joliot-Curie  supercomputing center thanks to a GENCI allocation under contract n$^{o}$A0090910016.
M.H.E.B. and D.J. are indebted to the CCIPL computational center installed in Nantes for (the always very) generous allocation of computational time.  M.W. thanks the RRZE for allocation of computational time and support, the ERASMUS+ traineeship program for their financial support and P.\,Hansmann as co-supervisor of his master's period.
\end{acknowledgments}



%

\end{document}